\documentclass[twocolumn]{aastex702}
\DeclareRobustCommand{\okina}{%
  \raisebox{\dimexpr\fontcharht\font`A-\height}{%
    \scalebox{0.8}{`}%
  }%
}

\usepackage{amsmath}
\newcommand{\OO}{\={O}tautahi--Oxford}

\shorttitle{Quasi-ISOs}
\shortauthors{Forbes et al.}

\graphicspath{{./}{figures/}}

\begin{document}

\title{There and back again: the quasi-interstellar objects}

\author[0000-0002-1975-4449]{John C. Forbes}
\affil{School of Physical and Chemical Sciences--Te Kura Mat\=u, University of Canterbury, Christchurch 8140, New Zealand}
\email{john.forbes@canterbury.ac.nz}

\author[0000-0003-3257-4490]{Michele T. Bannister}
\affil{School of Physical and Chemical Sciences--Te Kura Mat\=u, University of Canterbury, Christchurch 8140, New Zealand}
\email{michele.bannister@canterbury.ac.nz}

\author[0000-0001-5578-359X]{Chris Lintott}
\affil{Department of Physics, University of Oxford, Denys Wilkinson Building, Keble Road, Oxford, OX1 3RH, UK}
\email{chris.lintott@physics.ox.ac.uk}

\author[0000-0001-6314-873X]{Matthew J. Hopkins}
\affil{School of Physical and Chemical Sciences--Te Kura Mat\=u, University of Canterbury, Christchurch 8140, New Zealand}
\email{matthew.hopkins@canterbury.ac.nz}

%\correspondingauthor{John C. Forbes}
%\email{john.forbes@canterbury.ac.nz}

\begin{abstract}

A population of interstellar objects (ISOs) exist that originate from the Solar System, rather than from other stars.
Such a foreground could challenge straightforward analysis of the ISO sample expected to be gathered by upcoming sky surveys.
We assess whether small bodies unbound from the Solar System can experience dynamical evolution in the Galactic potential that places them on re-encounter trajectories.
We find that these `quasi-interstellar objects' (quasi-ISOs) primarily depart the Solar System through erosion of the outer Oort cloud in the past few hundred Myr, excluding the most recent $\sim10$ Myr.
After orbiting in the Milky Way potential nearby the Sun but beyond the tidal radius, those ejected on certain orbits can re-encounter the Solar System.
Meanwhile, the larger population of ISOs produced by the Solar System early in its life will be too spread-out in the Galaxy to contribute significantly to the observed sample. 
We predict that quasi-ISOs will be intrinsically rare and have $v_\infty$ values of order 0.1 km s$^{-1}$, easily distinguishable from ISOs from other stars, meaning that the observed ISO sample will be truly Galactic. 
The detection of a quasi-ISO would imply larger-than-expected losses from the Oort cloud, or a particularly catastrophic erosion event $10-300$ Myr ago that would not be detectable any other way.

\end{abstract}

%% https://astrothesaurus.org

\section{Introduction} \label{sec:intro}

Interstellar objects (ISOs), theorized for decades \citep[e.g.][]{sekanina1976,mcglynn1989,stern1990_planetformation,Jewitt_2003,Engelhardt_2017}, became an observational reality with the detection of 1I/\okina Oumuamua \citep{Meech_2017}. 
They are distinguished by their hyperbolic orbits with orbital energies unambiguously positive and well-separated from objects originating in the Oort cloud.
Soon after its discovery, the possibility of 1I instead being an inwardly scattered Oort cloud object was raised. 
This Solar System origin is disfavoured due to the unbinding and scattering dynamics relative to the current Oort cloud and the hyperbolic trajectory of 1I \citep{Higuchi_2020}. 
Note that this is distinct from the case of ISOs captured into the Solar System's small-body populations and resident for some tens of Myr \citep[e.g.][]{Hands_2020,Napier_2021_capture,Napier_2021_fate}; no such objects are known at present \citep{Morbidelli:2020}.

Interstellar objects are generally believe to have originated in earlier planetesimal disks \citep{'OumuamuaISSITeam_2019}, from which they are unbound either directly by interactions with planets \citep[e.g.][]{Moro-Martin_2018,albrow2026,huang2026,monk2026} or by external perturbations to the exo-Oort cloud of the planetary system \citep{Moro-Martin_2019b,portegieszwart2021}. Once ejected from its parent star system, an ISO may be delayed yet again by orbiting in the birth cluster of its parent star \citep{wu2023,wu2024,Forbes_2024,flamminidotti2025}, and in fact stellar encounters in the birth cluster can shape the ejection dynamics of ISOs \citep{torres2026}.
Eventually, after one, two, or all three of these processes, the ISO will escape to orbit independently in the Galaxy. 

The ISOs from any given progenitor will form a tidal stream \citep{portegieszwart2021, Forbes_2024} that stretches away from the progenitor. 
The stream expands at a speed of order of the larger of the following two velocities: the typical ejection velocity, and $\Omega r_\mathrm{Hill}$, the speed difference due to the Galaxy's differential rotation at the Hill sphere. Here $r_\mathrm{Hill} = (GM/(4\Omega^2-\kappa^2))^{1/3} \sim 1.5\ \mathrm{pc} (M/M_\odot)^{1/3}$ for objects in the Solar neighborhood, where $\kappa=\sqrt{2(\beta+1)}\Omega$ is the epicyclic frequency, and $\beta=d\ln v_\mathrm{circ}/d\ln r$ is the local powerlaw index of the rotation curve. 
The former typically applies to individual stars, while the latter applies to ejections from star clusters. 
The velocity dispersion of the stream will increase over time due to dynamical heating from inhomogeneities in the Galactic disk \citep[e.g.][]{kamdar2021,Forbes_2024}. 

The progenitor will remain near the maximum density of the ISO stream \citep{Forbes_2024}.
This suggests that it is plausible for a star system like the Sun to re-encounter the ISOs it has produced, at a non-negligible rate. 
We term these `quasi-interstellar objects', or `quasi-ISOs', in common with quasi-satellites.

If quasi-ISOs are abundant in the ISO sample that will shortly be discovered by new sky surveys \citep[e.g.][]{Dorsey_2025}, they raise the specter of a local foreground.
For instance, if they have dynamical properties in common with other ISOs, but are drawn from the compositional variation present within the Solar System (e.g. the carbon-depletion range seen in comets \citep{Opitom2026}), quasi-ISOs would lead to the assumption that the Galactic population of ISOs has great overlap with the Solar System small-body populations. Misidentification could also lead to dramatic errors in the inferred number density of ISOs.
This would create challenges for the interpretation of the Galactic ISO population and its many opportunities \citep[e.g.][]{Lintott_2022}. 
Indeed, ISOs would then be poorly suited to testing the Copernican principle as it applies to planetary formation.

In Section \ref{sec:prob}, we write down and integrate the equations of motion for a series of test-particle integrations, in which a burst of ISOs is produced by the Sun at some time in the past, and then integrated forward in the Galactic potential with a dynamical heating model. In Section \ref{sec:oort}, we apply several models of Oort cloud erosion to predict how the aforementioned burst models should be weighted to calculate the integrated rate of quasi-ISO encounters at the present day. In Section \ref{sec:prop} we show the quasi-ISO rate, their distribution of $v_\infty$, and their on-sky radiants. Implications for quasi-ISOs as a foreground population and for the potential discovery of quasi-ISOs are discussed in Section \ref{sec:disc}, and we conclude in Section \ref{sec:conc}.

\section{The probability that a Solar System ISO returns}
\label{sec:prob}
Our goal is to assess how interstellar objects that originate in the Solar System may travel the Galaxy and return to the Solar System at the present day. Our approach is to integrate the Sun's position and velocity backwards in time. We then run a sequence of simulations wherein a burst of interstellar objects is produced by the Solar System, which we trace with $N$ test particles. We integrate the motions of the Sun and the test particles forward, then at the epoch of the present day, we estimate the rate at which the Sun's ISO population represented by the test particles encounters the inner Solar System. These rates are of course proportional to the number of ISOs presumed to be ejected in each ejection event. We therefore need to convolve the results of these simulations with a model for how ISOs are ejected from the Solar System over time, which is the subject of Section \ref{sec:oort}. 

In each simulation, the ISOs are ejected from a 1.5 pc shell around the Sun's location at ejection time. The ISOs are placed on the 1.5 pc shell such that their initial velocity is purely in the radial direction away from the Sun. To shift to the Galaxy frame, we simply add the Sun's position and velocity vectors $\vec{x}_\odot$ and $\vec{v}_\odot$ at the time of ejection. The 3D velocities are drawn from a 3D normal distribution with an isotropic diagonal covariance matrix with elements $\sigma_\mathrm{ej}^2$ except as noted below. The magnitudes of these vectors are increased by adding in quadrature the escape velocity from the Sun at 1.5 pc to ensure that all particles ejected are formally unbound. Integrations are carried out with the \texttt{DifferentialEquations.jl} package in \texttt{julia} rather than our \texttt{lbparticles} package \citep{forbes2026} because of the importance of dynamical heating.

Each of the simulations has the following parameters
\begin{itemize}
    \item $t_\mathrm{burst}$, the time at which the ISOs are released. We use values ranging from $t_\mathrm{burst}=t_0 - 1\ \mathrm{Myr}$ to $t_\mathrm{burst}=t_0 - 800\ \mathrm{Myr}$, where $t_0$ is today.
    \item $\sigma_\mathrm{ej}$, the velocities the ISOs are given relative to the Sun. We use $\sigma_\mathrm{ej} = 0.1\ \mathrm{km}\ \mathrm{s}^{-1}$ or $\sigma_\mathrm{ej} = 1\ \mathrm{km}\ \mathrm{s}^{-1}$, appropriate for flybys ejecting Oort cloud objects. Larger velocities are possible from a variety of mechanisms \citep[e.g.][]{Pfalzner_2021,albrow2026} but we will argue that they likely do not contribute to the quasi-ISO population. 
    \item The heating model, one of ``white noise,'' ``correlated,'' or ``none.'' (see below).
    \item Latitude scaling. Typically ejections are isotropic, but we also run a case where the ejection distribution is flattened by reducing the $v_z$ component of the ejection velocity by a factor of 10. This is done to mimic the distribution of late-time ejections in the Solar System simulation of \citet{nesvorny2023}, visible in Figure 7 of \citet{albrow2026}. We believe their concentration of ejections near zero Galactic latitude is the result of the Galactic tide confining ejecta moving slower than $\sim \Omega r_\mathrm{Hill}$.
\end{itemize}

\subsection{Heating models}
 
\begin{figure*}
    \centering
    \includegraphics[width=\linewidth]{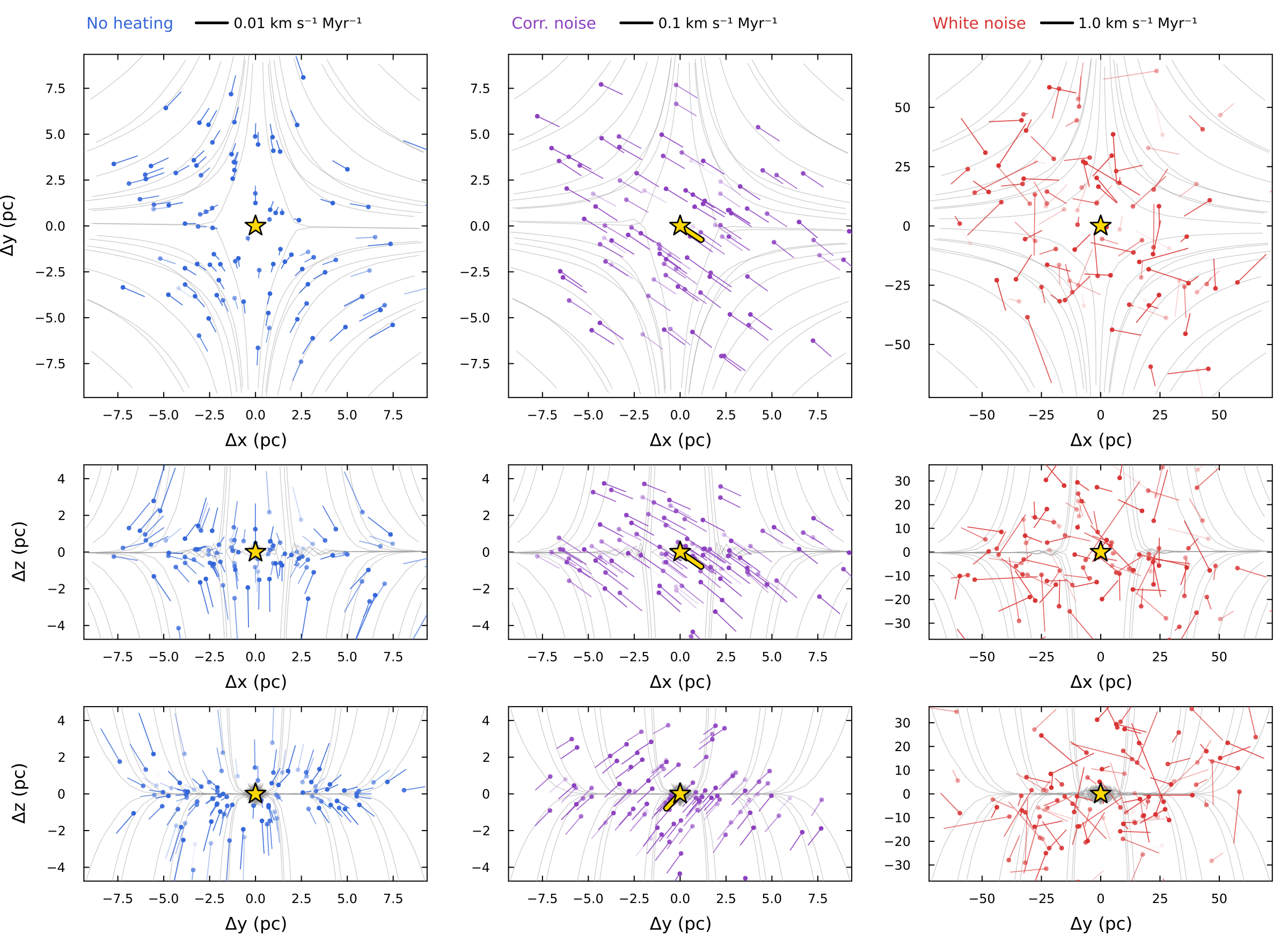}
    \caption{The dynamical heating models. Each panel shows 100 particles from a stream evolved under the given heating model over 30 Myr. Particles are shown in coordinates centered on the Sun (yellow star). The tails on each point show the negative direction of acceleration felt by the particle, relative to the Sun's acceleration from the Galaxy's background potential. The streamlines in each panel show the smooth potential's acceleration relative to the Sun's smooth acceleration, i.e. the tidal field excluding perturbations. Note the different scales of the acceleration lines in each column, and the larger spatial extent in the 3rd column. }
    \label{fig:heatingmodels}
\end{figure*}

In our integrations we test 2 simple models of dynamical heating (and another set of models with no heating, i.e. orbits in a smooth time-invariant potential). See Figure \ref{fig:heatingmodels}.

\subsubsection{White Noise}
The simplest heating prescription is to include a white noise term in the equations of motion as follows. While the positions are evolved as usual, with $dx_i/dt = v_i$ for each component, the velocities obey a stochastic differential equation
\begin{equation}
    dv_i = -\nabla \Phi_\mathrm{smooth}(\vec{x}) dt - G M_\odot (\vec{x}-\vec{x}_\odot)/|\vec{x}-\vec{x}_\odot|^3 + s~dW.
\end{equation}
Here $dW$ is a standard differential white noise term. In a time-invariant potential, included here as the first term, energy is conserved, so the dynamical heating enters only via the stochastic term. For the smooth term we use the MilkyWay2014 potential from \citet{bovy2015}. An ensemble of particles evolved under this equation would develop a local velocity dispersion of approximately $\sigma = s \sqrt{t}$ after time $t$. We can therefore set $s = (35\ \mathrm{km}\ \mathrm{s}^{-1}) / \sqrt{14\ \mathrm{Gyr}}$ to model a constant heating rate ($s^2$) that accounts for the majority of the observed age-velocity dispersion relation (AVR) in local stars \citep{holmberg2009}, though the usual caveats apply, namely some component of the observed AVR may be due to higher stellar birth velocity dispersions in the past \citep{bournaud2009,forbes2012}.

\subsubsection{Correlated Noise}
While the white noise prescription may be used to produce a reasonable increase in energy, it lacks correlations in space and time that might be expected for realistic heating processes. For example, if heating arises from encounters with giant molecular clouds, ISOs near each other should experience very similar velocity kicks, and those kicks should not change by a large amount from one Myr to the next. 

To account for the correlation in time, we employ a damped random walk (Ornstein-Uhlenbech, or OU) process \citep{uhlenbeck1930} which obeys
\begin{equation}
\label{eq:ou}
    du_i = -u_i \tau^{-1} dt + S dW,
\end{equation}
where the $u_i$ will be used as perturbations to the velocity,
\begin{equation}
\label{eq:oueom}
    dv_i/dt = -\nabla \Phi_\mathrm{smooth}(\vec{x}) - G M_\odot \frac{\vec{x}-\vec{x}_\odot}{|\vec{x}-\vec{x}_\odot|^3} + \alpha u_i + \beta u_{i,\odot}.
\end{equation}
The damped random walk has a characteristic correlation timescale of $\tau$; the variance per unit frequency (the power spectral density) on frequencies ($f$) higher than $\tau^{-1}$ (timescales shorter than $\tau$) fall as $f^{-2}$, while on longer timescales the variance per unit frequency is flat (white noise). We choose a value of $S$ that accounts for the full AVR, as we did with the white noise case, so that $S=s/\sqrt{\tau}$.

To account for spatial correlations, we have also introduced mixing factors $\alpha$ and $\beta$, which are between 0 and 1 but not necessarily constant. It enforces a correlation between the kicks experienced by the Sun, $u_{i,\odot}$ and the kicks experienced by each ISO, $\alpha u_i + \beta u_{i,\odot}$. The correlation between these two velocity kick time series is $\beta/\sqrt{\alpha^2 + \beta^2}$, so we set $\alpha^2 = 1-\beta^2$, and 
\begin{equation}
    \beta = \exp\left(-\frac12 \frac{\left|\vec{x}-\vec{x}_\odot\right|^2}{r_\mathrm{corr}^2}\right),
\end{equation}
which sets the correlation between the Sun's heating and the ISOs' heating to $\beta$ identically. The Sun's heating, $u_{i,\odot}$ is evaluated through Equation \ref{eq:ou} and its motion is evaluated through Equation \ref{eq:oueom} with $\beta=0$, that is
\begin{eqnarray}
        du_{i,\odot} &=& -u_{i,\odot} \tau^{-1} dt + S dW \nonumber \\
    dv_{i,\odot}/dt &=& -\nabla \Phi_\mathrm{smooth}(\vec{x}_\odot) + u_{i,\odot}.
\end{eqnarray}
Essentially the Sun experiences kicks that are correlated in time on some timescale $\tau$, then the ISOs experience kicks that are strongly correlated with the Sun's when the separation between the Sun and the ISO is less than the correlation length, $r_\mathrm{corr}$. This treatment does neglect additional correlation between nearby pairs of ISOs when they are far from the Sun. This is done mostly for practical convenience, to keep the ISO integrations easily parallelizable. It is also a reasonable approximation because, as we will see, most quasi-ISOs never get particularly far from the Sun anyway.

In our integrations, we set $\tau = 17$ Myr, and $r_\mathrm{corr} = 170$ pc. These values correspond to the correlation timescale and distance scale if a particle on our assumed Sun-like orbit encounters giant molecular clouds randomly distributed throughout the Solar circle, such that their total surface density corresponds to the observed surface density of molecular hydrogen in the Solar circle. This is an approximation to the full density structure of the interstellar medium \citep{modak2026}, and neglects spiral arms, in which molecular clouds are preferentially found \citep[e.g.][]{heyer2015}.

\subsubsection{Backwards Integration of the Sun}
A key step regardless of the heating model is the backwards integration of the Sun. In a smooth time-invariant potential, i.e. the no heating model, integrating the equations of motion backwards and then forwards yields the initial starting point up to finite time step errors introduced by the integrator. In the context of stochastic differential equations, this is not true - integrating backwards adds noise, just like integrating forwards. In the white noise heating model, we address this by not including the noise when we integrate the Sun's trajectory; only the ISOs experience heating. This ensures that the Sun does not random walk out of the densest concentration of ISOs, and avoids the backward- forward- integration asymmetry. This introduces some regularity into the white noise model beyond what one would get from treating all particles in the simulation as truly independently subject to white noise velocity kicks.

In the correlated heating case, we do include random kicks to the Sun's velocities, since these are also applied to nearby ISOs. The Sun's OU process, the $u_{i,\odot}$, is simulated first and cached. Once these are fixed, we have the option of integrating backwards from the Sun's current position and velocity, which would guarantee the Sun arrives at its current position and velocity, again up to integrator error. However, this will tend to result in the nonphysical situation that the Sun gains random energy over the course of the backwards integration. To avoid this possibility, we place a prior on the OU noise as follows. When the Sun's current position is integrated backwards subject to the OU noise, we evaluate the Sun's non-circular energy 4.5 Gyr ago, namely $E_\mathrm{non-circ} = (1/2)(v_r^2 + \kappa^2(R-R_g)^2 + v_z^2 + \nu^2z^2)$, where $R_g$ is the guiding center radius of the orbit (where $R_g v_\mathrm{circ}(R_g) = R v_\phi$). We perform this integration for 64 draws of the OU process, and select the one that yields a value of $E_\mathrm{non-circ}$ in the $\approx$10th percentile.  Essentially integrating the Sun backwards yields a wide variety of solutions that are consistent with the Sun's current position and velocity, but physically we expect that due to dynamical heating, the Sun, all else equal, should have been dynamically colder in the past. Note that our results are insensitive to this choice, since to a first approximation, all that matters is the relative separation between the Sun and the ISOs.

\subsection{The Encounter Rate for Each Simulation}
Integrating the encounter rate of the ISOs with the Solar System requires some care. Not only is the cross-section of the encounter velocity-dependent, but the {\em direction} of the motion of the particles matters. Early in each simulation, the particles are close to their initial values, moving radially away from the Sun on unbound orbits. Typical integrals of the velocity distribution at a fixed point \citep[e.g.][]{Forbes_2019, hopkins2025a} would not distinguish between ingoing and outgoing particles, and would therefore erroneously give a non-zero encounter rate in these cases.

We therefore evaluate the rate on a finite sphere centered on the point of interest (the Sun), and restrict contributions to the rate to incoming particles, 
\begin{eqnarray}
    \mathcal{R} = r^2 &\int_{-1}^1 d\cos\theta\int_0^{2\pi} d\phi\int_0^\infty dv \Big( \nonumber \\
    &\int_{\cos\theta_c}^1 d\cos\theta_v \int d\phi_v\ v^2 (\vec{v} \cdot \hat{r}) f \Big).
\end{eqnarray}
Here $r$, $\theta$, and $\phi$ are spherical coordinates centered on the Sun, and $v$, $\theta_v$ and $\phi_v$ are spherical coordinates in velocity space aligned around $\hat{r}$, so that the $\hat{r}$ direction at a given point on the sphere corresponds to $\theta_v=\pi$, and $\theta_v=0$ points directly towards the Sun. 
The integral is therefore simply the flux of ISOs through the sphere, integrated over the velocity distribution at each point on its surface. 
The velocity integral is restricted to values of $\theta<\theta_c$, where $\theta_c$ is the maximum angle at which an ISO can be moving such that its pericenter $q<q_\mathrm{max}$. The density being integrated is $f$, the phase space density of ISOs. We normalize $f$ so that its integral over all six dimensions of space and velocity is 1.
 
Because we can typically assume $\theta_c\ll 1$, we neglect the small variations in the velocity part of $f$ over this small range of $\theta_v$, yielding
\begin{eqnarray}
    \mathcal{R}\approx& 2\pi r^2 \int_{-1}^1 d\cos\theta\int_0^{2\pi} d\phi \Big( \quad \quad \nonumber \\
   &  \int_0^\infty dv\ v^3 f \int_{\cos\theta_c}^1 d\cos\theta_v\ \cos\theta_v \Big),
\end{eqnarray}
where we have also rewritten the dot product in terms of $\theta_v$. The innermost integral can now be readily evaluated. We note that $\theta_c$ does depend on $v$, so 
\begin{equation}
    \mathcal{R}\approx \pi r^2 \int_{-1}^1 d\cos\theta\int_0^{2\pi} d\phi\int_0^\infty dv\ v^3 f  \sin^2\theta_c 
\end{equation}

We then evaluate this integral via importance sampling with a Gaussian proposal distribution $Q(\vec{v})\sim N(\vec{\mu},\Sigma)$, where $\mu$ is the sample mean velocity of the $K=300$ nearest neighbors in physical space, and $\Sigma$ is four times the sample covariance of those same particles, with the diagonal components increased by an additional 5\%. Then
\begin{equation}
    \mathcal{R} \approx \pi r^2 N^{-1} \sum_{i=1}^N v_i f_i \sin^2 \theta_{c,i} \mathcal{I}(v_i>v_{\mathrm{esc},i}) / Q(\vec{v}_i)
\end{equation}
For each sample $i$ we draw a velocity vector $\vec{v}_i$ from the proposal distribution $Q(\vec{v})$, then evaluate $f$ at velocity $\vec{v}_i$ and corresponding position $\vec{x} = \vec{x}_\odot - r\ \vec{v}_i/v_i$.

Finally, $\theta_c$ itself can be obtained by noting that the angular momentum of the particle relative to the Sun is $r v_\mathrm{tan}$, where $v_\mathrm{tan}$ is the component of the velocity perpendicular to the surface of the sphere. For a hyperbolic orbit, the angular momentum is also equal to $b v_\infty$, where $b$ is the impact parameter and $v_\infty$ is the velocity of the particle in the limit that its velocity has not yet been influenced by the Sun, namely $v_\infty^2 = v^2 + v_\mathrm{esc}^2$, and $v_\mathrm{esc}^2 = 2 G M_\odot/r$. The impact parameter and the particle's pericenter, $q$, are related via
\begin{equation}
    b = q\sqrt{1+2 G M_\odot/(q v_\infty^2)}
\end{equation}
By conservation of angular momentum, $v_\mathrm{tan}/v_\infty = b/r$, so to find the critical value of $\theta_v$, we set
\begin{eqnarray}
    \sin\theta_c &=& \frac{v_\mathrm{tan}}{v} = \frac{v_\mathrm{tan}}{v_\infty \sqrt{1 - 2GM_\odot/(r v_\infty^2)}} \nonumber \\
    &=& \frac{q}{r}\sqrt{\frac{1+2 G M_\odot/(q v_\infty^2)}{1 - 2GM_\odot/(r v_\infty^2)}}
\end{eqnarray}

\subsection{Results}

\begin{figure*}
    \centering
    \includegraphics[width=\linewidth]{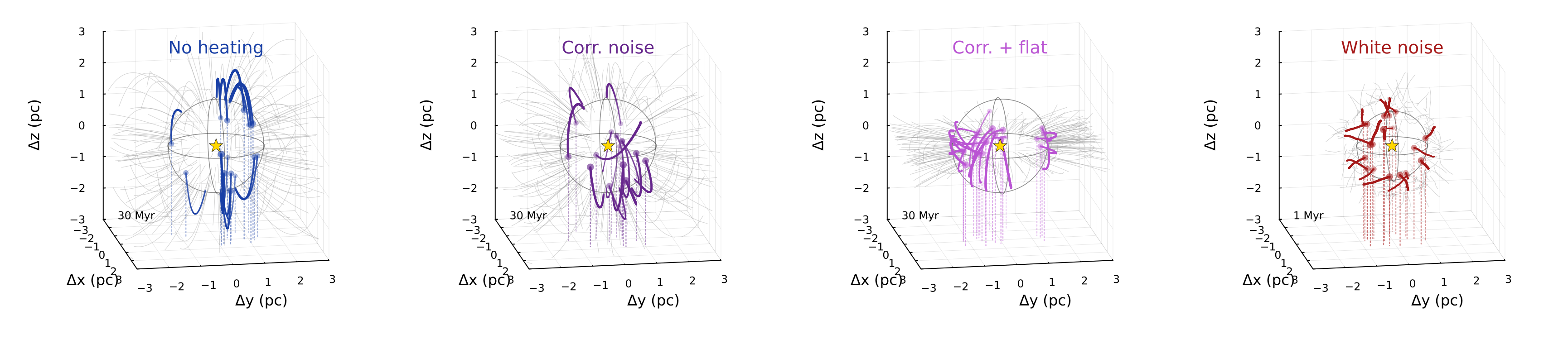}
    \caption{Quasi-ISO trajectories. Individual trajectories are shown in a frame comoving with the Sun (located at the origin). ISOs that are ejected from the Solar System and return (to within 1.5 pc) are shown in color, while the gray lines show ISOs that do not return. Distance to the camera is encoded in the thickness and transparency of the lines, and the terminal points where the particles re-enter the 1.5 pc sphere are shown with circular markers. The markers' locations are projected down to the plane at the bottom of the plot at $\Delta z=3\ \mathrm{pc}$. Note that in the case of white noise, the trajectories are shown over a much shorter time (1 Myr) since by 30 Myr very few ISOs in this scenario remain close to the Sun.}
    \label{fig:traj3d}
\end{figure*}

For the simulations that produce a burst of ISOs 30 Myr in the past\footnote{This time corresponds to peaks in the most realistic heating models, which is the result of the vertical oscillation period of the disk (see below).}, we save the full trajectories of the particles, a subset of which we show in Figure \ref{fig:traj3d}. 
The particles that return to the Sun's sphere of influence in that time frame show distinct patterns.
Most notably, in the absence of white noise-style heating, particles ejected with substantial velocities in the $y-$direction preferentially do not return. 
In these cases, the vertical motion of the particles, if it is allowed by the initial conditions, contributes a substantial number of returning particles. 
The $y-$direction corresponds to the tangential component of the Sun's motion around the Galaxy, in other words the direction in which the nascent stream of ISOs will spread out. 
The vertical and radial components, meanwhile, will return to the sphere of influence via their epicyclic motion. 
We will return to this point in section \ref{sec:prop}, where we show sky maps of all quasi-ISOs integrated over the full modelled loss histories of the Oort cloud.

From each simulation, we evaluate the encounter rate as described in the previous section on a $0.1$ pc sphere centered on the Sun at the present epoch. We set the maximum pericenter distance to $q_\mathrm{max} = 5$ au. Each simulation's phase space density is estimated via an adaptive cross-validated kernel density estimator \citep{Forbes_2024}. In addition to estimating the encounter rate, the importance-sampled weights allow us to predict, for each simulation, a distribution of excess velocities, $v_\infty$, and incoming radiants $(\theta,\phi)$.

\begin{figure*}
    \centering
    \includegraphics[width=1\linewidth]{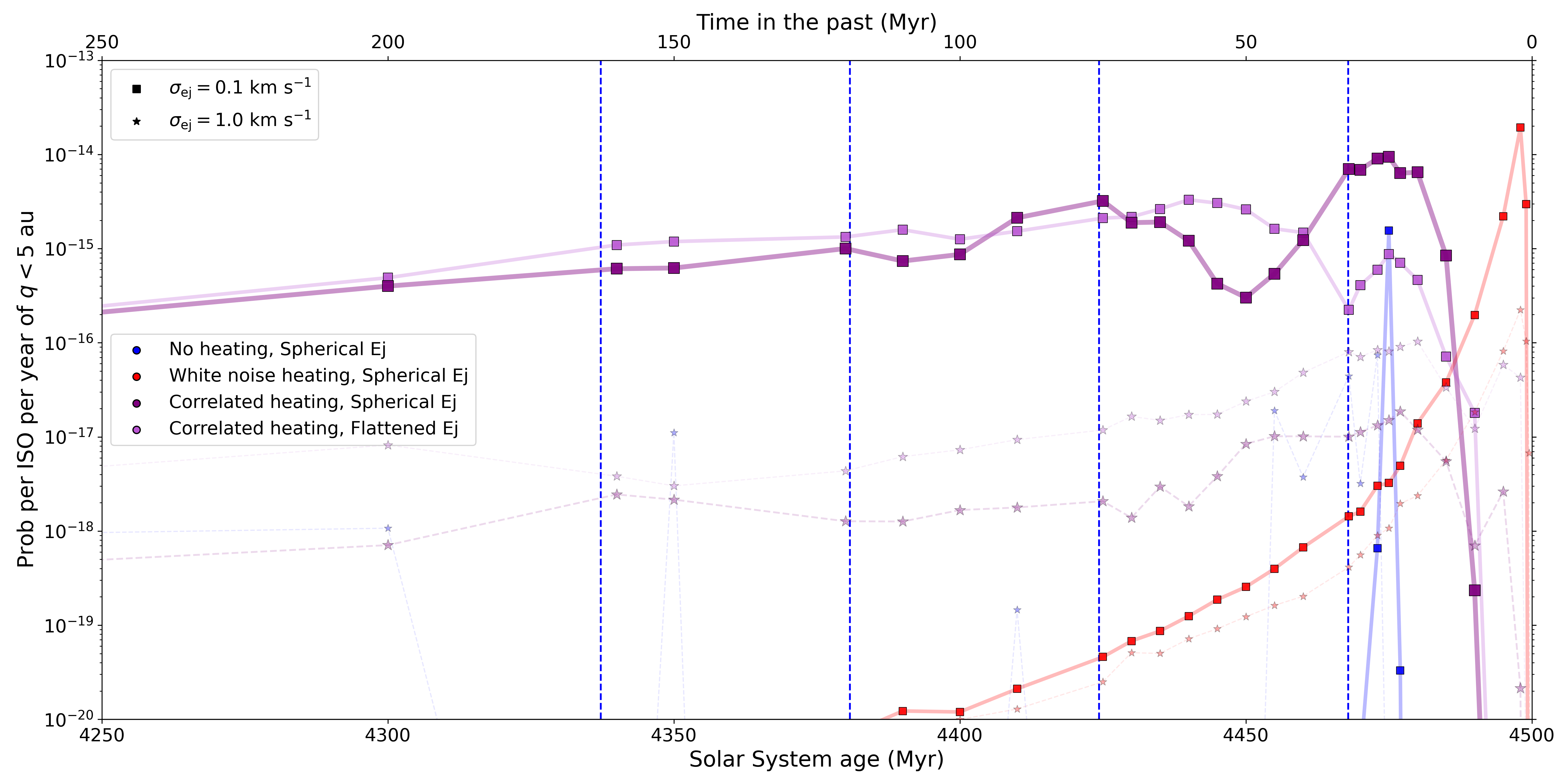}
    \caption{The re-encounter probability. For each single-burst simulation, we compute the probability per year at the present day that a random ISO will re-encounter the Solar System with a pericenter less than 5 au. The bold lines show the $\sigma_\mathrm{ej}=0.1\ \mathrm{km}\ \mathrm{s}^{-1}$ case, which is almost always much higher than the faster-ejection case of $\sigma_\mathrm{ej}=1\ \mathrm{km}\ \mathrm{s}^{-1}.$ White noise heating produces encounters only when the ejection event was within the past few Myr, whereas other heating models only produce quasi-ISOs on 10s-100s of Myr timescales.} 
    \label{fig:probs}
\end{figure*}

In Figure \ref{fig:probs}, we show the encounter rates from all of the burst simulations. Each heating or initial conditions model is shown in a different color, with the $\sigma_\mathrm{ej}=0.1\ \mathrm{km}\ \mathrm{s}^{-1}$ case emphasized with bolder lines. Each point is normalized so that the phase space density integrated over all space and time is $1$, so the resulting rates should be interpreted as the probability per year of  a re-encounter for a randomly-selected object in the ejection burst. Notably no scenario produces a probability much greater than $10^{-14}\ \mathrm{yr}^{-1}$ per ISO.

The primary result is that large velocity dispersions or white noise heating can produce prompt re-encounters, where the object re-encounters the inner Solar system within $\sim 1$ Myr of being ejected. For the more realistic cases, the Sun's ISOs do not tend to return for another 30-100 Myr, comparable to the vertical and epicyclic oscillation periods in the Solar neighborhood of the Galaxy. 

The dashed vertical lines show the times when re-encounters are most likely if they are being driven by vertical simple harmonic motion governed by
\begin{equation}
    \ddot{z} +\nu^2 z = 0.
\end{equation}
Because this is a linear equation, the perturbation in $z$ relative to the Sun, $\Delta z = z-z_\odot$ obeys the same equation,
\begin{equation}
    d^2{\Delta z}/dt^2 + \nu^2 \Delta z =0.
\end{equation}
If each particle is initialized with $\Delta z = \delta z$ and $v_{\Delta z} = d\Delta z/dt = \delta v$, the initial value problem has the solution
\begin{equation}
    \Delta z(t-t_\mathrm{burst}) = \delta z \cos(\nu (t-t_\mathrm{burst})) + \frac{\delta v}{\nu} \sin(\nu (t-t_\mathrm{burst})).
\end{equation}
Solving for $t$ when $\Delta z =0$, we find
\begin{equation}
    t_0 - t_\mathrm{burst} = \nu^{-1}( n\pi - \arctan(1/\alpha) ),
\end{equation}
where $\alpha=\delta v/(\nu \delta z)=$ const. for spherical ejections, and $n$ is any integer. Note that for these vertical lines we have assumed $\nu=\mathrm{const.}$, which is only true for circular orbits with small vertical oscillation amplitudes, neither of which strictly applies here, so only the first few solutions for $\Delta z = 0$ are plotted. Both the no heating case (blue) and the correlated heating case (dark purple) show marked peaks near the first two of these solutions, and a marked deficit between them. When the vertical oscillations are essentially removed by assuming that objects are preferentially ejected in the Galactic midplane (light purple), these features do not appear.

As $t_0-t_\mathrm{burst}$ gets large (the burst occurs further in the past; right-hand side of the left panel of Figure \ref{fig:probs}), the probability of re-encountering the Sun plummets regardless of the model. This is because the stream of interstellar objects produced has a characteristic density of order $\kappa\nu\sigma^{-3}t^{-1}$ \citep{Forbes_2024}, so as $\sigma$ increases\footnote{For the no heating case, the local velocity dispersion increases as well despite the lack of explicit heating due to the particles in the stream experiencing varying vertical oscillation frequencies \citep{dehnen2018}}, the density of the stream decreases. Therefore even though the vast majority of planetesimals ejected by the Solar system were likely lost at early times, of order 10 Myr after formation, due to giant planet scattering, these ISOs contribute almost nothing to the quasi-ISO population.

\section{Building the Total Encounter Rate from Oort Cloud Losses}
\label{sec:oort}

Given a specific choice of ejection model ($\sigma_\mathrm{ej}$ and geometry), and heating model, we can use our simulation results at a sequence of $t_\mathrm{burst}$ values to estimate the total rate of quasi-ISOs entering the inner Solar System as
\begin{equation}
\label{eq:rtot}
    \mathcal{R} = \int_0^{t_0} \tilde{\mathcal{R}}_\mathrm{burst}(t) \frac{dN_\mathrm{ej}}{dt} dt
\end{equation}
where $dN_\mathrm{ej}/dt$ is the rate at which the Oort cloud is losing objects as a function of time, and $\tilde{\mathcal{R}}_\mathrm{burst}$ is the probability of re-encounter per year shown in Figure \ref{fig:probs}. 
Equation \ref{eq:rtot} may be understood as a reweighting of the ejection rate (the second factor) by the probability per unit time that such an ejection will result in a quasi-ISO (the first factor). 
In general because we have only done simulations at a finite set of $t_\mathrm{burst}$ values, we use a linear interpolation of $\log \tilde{\mathcal{R}}$ with respect to $\log t$ to evaluate Equation \ref{eq:rtot}, i.e. exactly the curves drawn between the points in the left-hand panel of Figure \ref{fig:probs}. 

The rate at which the Oort cloud loses comets is highly uncertain. 
In the literature, typical timescales for the Oort cloud to lose of order its mass range from 3 to 13 Gyr \citep{portegieszwart2025}. 
Because the Galactic tide rotates the orbital elements of Oort cloud comets \citep{heisler1986}, and it is particular parts of phase space that are susceptible to being lost at any given moment, one generally assumes that some fraction $f$ of the Oort cloud is lost due to any particular event or at any particular time. 
This means that the number of objects (neglecting new entries into the Oort cloud) and the rate at which objects are lost should both be exponential, e.g. $N_\mathrm{Oort} \propto dN_\mathrm{ej}/dt \propto \exp{(-t/t_\mathrm{loss})}$. 
The slowest loss rates correspond to losses from the influence of the Galactic tide, which deforms the equipotential surface around the Sun over the course of the Sun's orbit. 
The potentially far larger source of Oort cloud erosion is stellar flybys, which may reasonably be approximated as impulsive events that typically have $f\ll 1$. 
The events that remove the largest fraction of the Oort cloud are rare encounters where the relative velocity between the perturbing object and the Sun are slow, $\lesssim 1\ \mathrm{km}\ \mathrm{s}^{-1}$ \citep{hanse2018}. 

We employ three different models for how the Oort cloud is eroded, i.e. $dN_\mathrm{ej}/dt$. 
The first is a simple exponential, where we simply assume that 
\begin{equation}
    \frac{N_\mathrm{Oort}(t)}{N_\mathrm{Oort}(0)} = \exp(-t/t_\mathrm{loss})
\end{equation}
with values of $t_\mathrm{loss}$ ranging from 3-13 Gyr. 
This is a useful smooth baseline.

We next consider the rate at which planetesimals are lost from the \citet{nesvorny2023} simulation. When planetesimals cross a $5\times 10^5$ au sphere centered on the Sun, their positions and velocities are recorded along with the epoch. We assume that objects ejected after $100$ Myr were part of the Oort cloud for the purposes of estimating $N_\mathrm{Oort}(0)$, then simply attribute all subsequent losses in that simulation to Oort cloud erosion for the purposes of estimating $d N_\mathrm{ej}/dt$. 
The timing of these ejections can then be used to construct the total rate, just as with any other erosion model (see next paragraph). 
We {\em also} use the \citet{nesvorny2023} simulation to initialize each individual particle ejected, including its position and velocity, run that simulation forward in time subject to the correlated heating model, and finally evaluate the rate directly from the final positions and velocities of the particles. 
In the latter case, we only simulate particles from the final 100 Myr of the simulation, because the KDE struggles to handle multiple populations of ISOs at different dynamical temperatures. 
These two cases are respectively ``N23 (ejection timing only)'' and ``N23 (last 100 Myr + kin)''.

Finally, we use a Monte Carlo method to directly estimate the influence of flybys. In this case $dN_\mathrm{ej}/dt$ may be understood as essentially a series of delta functions at the times of the flybys, yielding 
\begin{equation}
\label{eq:rtotprod}
    \mathcal{R} \approx N_\mathrm{Oort}(0) \cdot \sum_i \left[  \tilde{\mathcal{R}}(t_i) F_i \left( \prod_{j\ \mathrm{s.t.}\ t_j<t_i} (1-F_j) \right) \right]
\end{equation}
where each stellar flyby is indexed by $i$, occurs at time $t_i$, and causes the loss of a fraction $F_i$ of the Oort cloud. The product accounts for all previous flybys up to flyby $i$. The overall rate is proportional to the (unknown) number of Oort cloud objects at the birth of the Solar System (or more accurately, shortly afterwards when the Oort cloud was in place), $N_\mathrm{Oort}(t=0)$. We also have that, in this simplified picture,
\begin{equation}
N_\mathrm{Oort} (0) \cdot  \left(\prod_i (1-F_i) \right)= N_\mathrm{Oort} (t_0).
\end{equation}
The loss fractions are found using the powerlaw fit from \citet{hanse2018},
\begin{eqnarray}
    F =\min&\Bigg(& 4.47\times10^7 \left(\frac{M_\mathrm{pert}}{M_\odot}\right)^{1.84} \left(\frac{b}{\mathrm{au}}\right)^{-2.17}  \nonumber \\
    &\times& \left(\frac{v}{\mathrm{km}\ \mathrm{s}^{-1}}\right)^{-1.91} 10^{\sigma_H x},\ 1\Bigg)
\end{eqnarray}
Each encounter contributes more as the perturber is more massive, closer, or slower, up to a fraction lost of 100\%. We also include an ad-hoc scatter $\sigma_H$ in dex to account for imperfections in the fit, which will increase the mean of $F$ (disregarding the clipping at 1) by a factor of $10^{\sigma_H^2 \ln(10)/2}$.

We then simply need to draw from an appropriate joint distribution of $M_\mathrm{pert}$, $b$, $v_\mathrm{rel}$, and $t_\mathrm{enc}$. To do so we follow the formalism of \citet{garciasanchez2001} updated by \citet{rickman2008}. Stars are divided up by spectral class (B0, A0, A5, F0, F5, G0, G5, K0, K5, M0, M5, white dwarfs, and giants), each of which has a velocity dispersion, mass range, number density, and mean velocity with respect to the Sun. We adjust the density of the M dwarf components so that the encounter rate from M dwarfs matches the Gaia DR2 completeness-corrected encounter frequency from \citet{bailerjones2018}. Encounters are assumed drawn from a uniform distribution over the Solar system's history. Masses are drawn uniformly from the mass range of each stellar type, and velocities are drawn from isotropic dispersion for each component to which the Sun's relative motion is added. Encounters are then weighted by the magnitude of the velocity and redrawn into an equal-weight sample using the \citet{hol2006} algorithm as implemented in \texttt{dynesty} \citep{speagle2020}. This step accounts for the fact that all else equal higher velocity encounters occur more frequently in proportion to the magnitude of the velocity. Finally, encounter impact parameters are drawn from $p(b) \propto \sqrt{b}$ between 0 and 2 pc. The many simplifications made here, including neglecting the Sun's current orbit, neglecting the possibility that the Sun migrated substantially during its history \citep{kaib2011,torres2019}, and neglecting the anisotropy in the velocity dispersion of each component \citep{holmberg2009}, are all subdominant relative to uncertainties in the \citet{hanse2018} formula, which itself is subdominant to uncertainties in both the distribution of the Oort cloud's orbital elements over time, and the synergistic role played by the Galactic tide alongside flybys in ejecting objects \citep{rickman2008}.

\begin{figure}
    \centering
    \includegraphics[width=\linewidth]{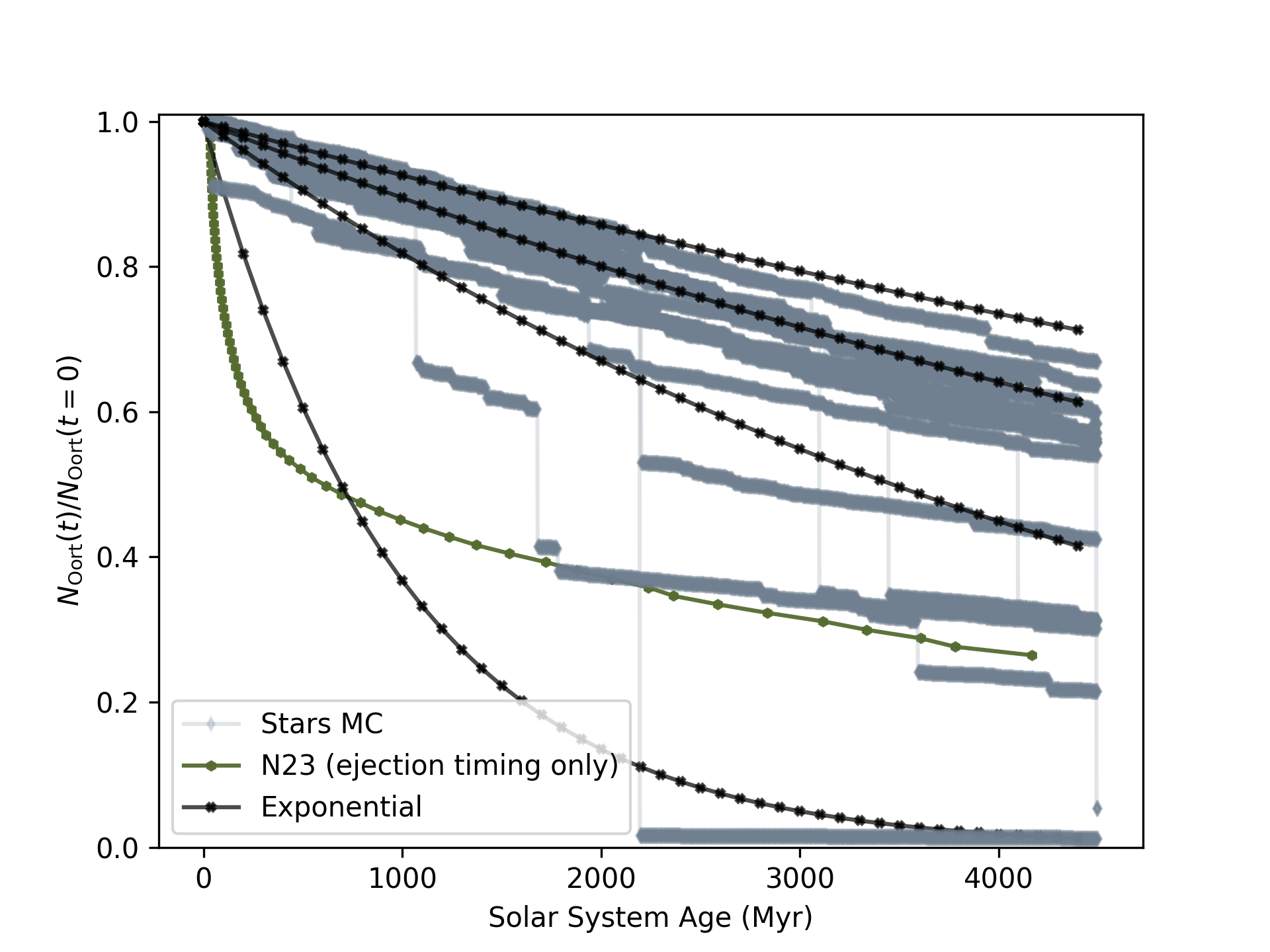}
    \caption{Models of Oort cloud erosion. The number of objects in the Oort cloud normalized to its value near the birth of the Solar system. N23 refers to the \citet{nesvorny2023} simulation.}
    \label{fig:erosion_models}
\end{figure}

The Oort cloud erosion histories produced by these three models --- simple exponentials, the \citet{nesvorny2023} N-body simulation, and the Monte Carlo stellar flyby approach --- are shown in Figure \ref{fig:erosion_models}. 
The scatter applied to the flyby loss formula is chosen to be $\sigma_H=1.1$ dex, which brings the stellar Monte Carlo approach into rough agreement with the other models on the fraction of the Oort cloud lost over 4.5 Gyr.

Finally, there is a question of normalization. We are interested not just in the fraction of the Oort cloud that returns to the Solar System after being ejected, but literally how many quasi-ISOs enter the observable volume per year. For any given pair of (erosion history, ejection and dynamics model), the rate of quasi-ISOs appearing within 5 au will be directly proportional to the overall normalization. 

One strategy to set the normalization is to measure the rate of dynamically new Oort cloud comets entering the inner Solar System, and carefully correct for the observational biases in their discovery. 
Using this method, \citet{boe2019} estimate there are about $5\times 10^{12}$ Oort cloud objects the size of 1I/\okina Oumuamua or larger present in the Oort cloud today, with order unity uncertainties, primarily from the uncertainty in identifying what fraction of long-period comets are dynamically new. 

Another strategy is to use the inferred number density of interstellar objects from observations within the Solar System, yielding $\sim 10^{16}$ objects larger than 1I from each star \citep{Raymond_2018a, Do_2018}.
If $\sim 95\%$ of planetesimals are ejected from the Solar System during its formation \citep{duncan1987AJ, dones2004, nesvorny2023}, then the Oort cloud should start with $\sim5\times10^{15}$ objects of this size. 
These two numbers are only consistent if the Oort cloud has lost $\sim 99\%$ of its objects over the lifetime of the Solar System, which introduces some tension between the two different normalization options. 
We will return to this issue in Section \ref{sec:discnorm}, but for now we will simply show results for both cases.

\begin{figure*}
    \centering
    \includegraphics[width=0.75\linewidth]{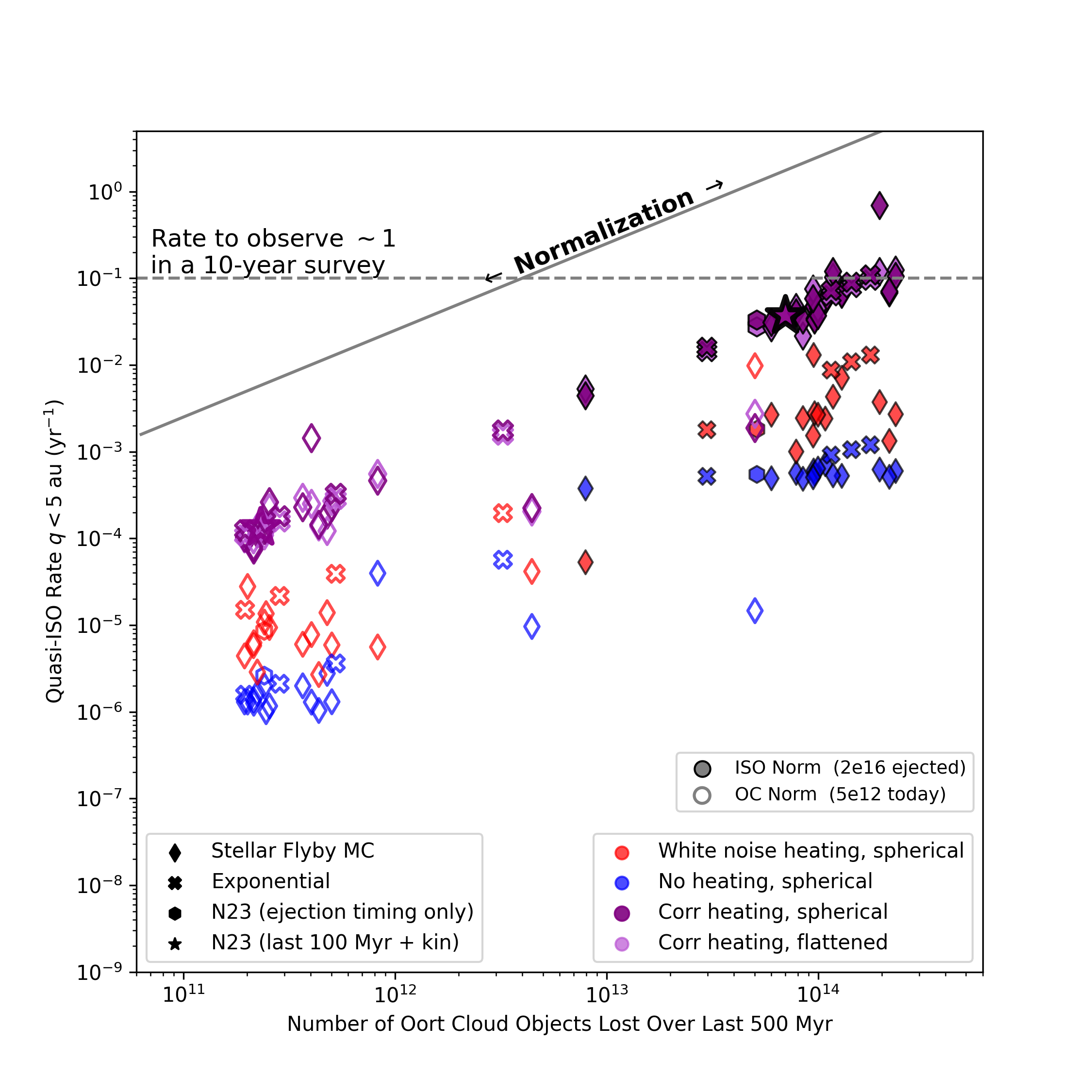}
    \caption{Predicted quasi-ISO rates. The y-axis shows our predictions for the present-day rate of quasi-ISOs entering the inner Solar system (pericenters $<$ 5 au), compared against the number of Oort cloud objects ejected over the past 500 Myr. Each set of colored points corresponds to a different heating+ejection model, while the different symbols correspond to Oort cloud erosion models. N23 refers to the simulation results from \citet{nesvorny2023}. In the case of ``ejection timing only'' we use the times that particles from the simulation cross the 500k au sphere in the simulation to interpolate the results of our burst simulations. In ``last 100 Myr + kin'' we initialize each particle with its position and velocity at the given time, integrate it, and evaluate the rate based on the resulting particle distribution. The filled vs hollow points correspond to different assumptions about the normalization of the Oort cloud, respectively one based on the ISO density, and one based on the rate of long-period comets. Changes in normalization move points parallel to the solid line (increasing the number of objects ejected directly increases the rate of quasi-ISOs). The rate required to expect one detection in a 10-year survey is of order $10^{-1}\ \mathrm{yr}^{-1}$, neglecting observational efficiencies.}
    \label{fig:erosion_result}
\end{figure*}

Figure \ref{fig:erosion_result} shows results normalized by the present-day number of Oort cloud objects in light colors on the left of the plot, and normalized by the inferred ISO density on the right in darker colors. The y-axis is the total rate integrated over the probability of each ISO returning times the rate at which ISOs are being ejected (Equation \ref{eq:rtot}), and the x-axis is the number of objects lost from the Oort cloud in the past 0.5 Gyr. As expected, the normalization choice moves the predictions along a 1:1 line, with the ISO-based normalization yielding higher quasi-ISO rates. The most realistic models involving correlated heating, and further, ejections flattened along the plane of the Galaxy, yield the highest quasi-ISO rates, potentially large enough to be detectable. However, the models that yield combined rates $\gtrsim 10^{-2}\ \mathrm{yr}^{-1}$ all require at least $5 \times 10^{12}$ objects to have been lost by the Oort cloud in the past 500 Myr. This is within the realm of reason, but is large, since it is comparable to the estimated number of Oort cloud objects present in the cloud today \citep{boe2019}.

\section{Dynamical Properties of the Quasi-ISOs}
\label{sec:prop}

\begin{figure*}[t]
    \centering
    \includegraphics[width=0.75\linewidth]{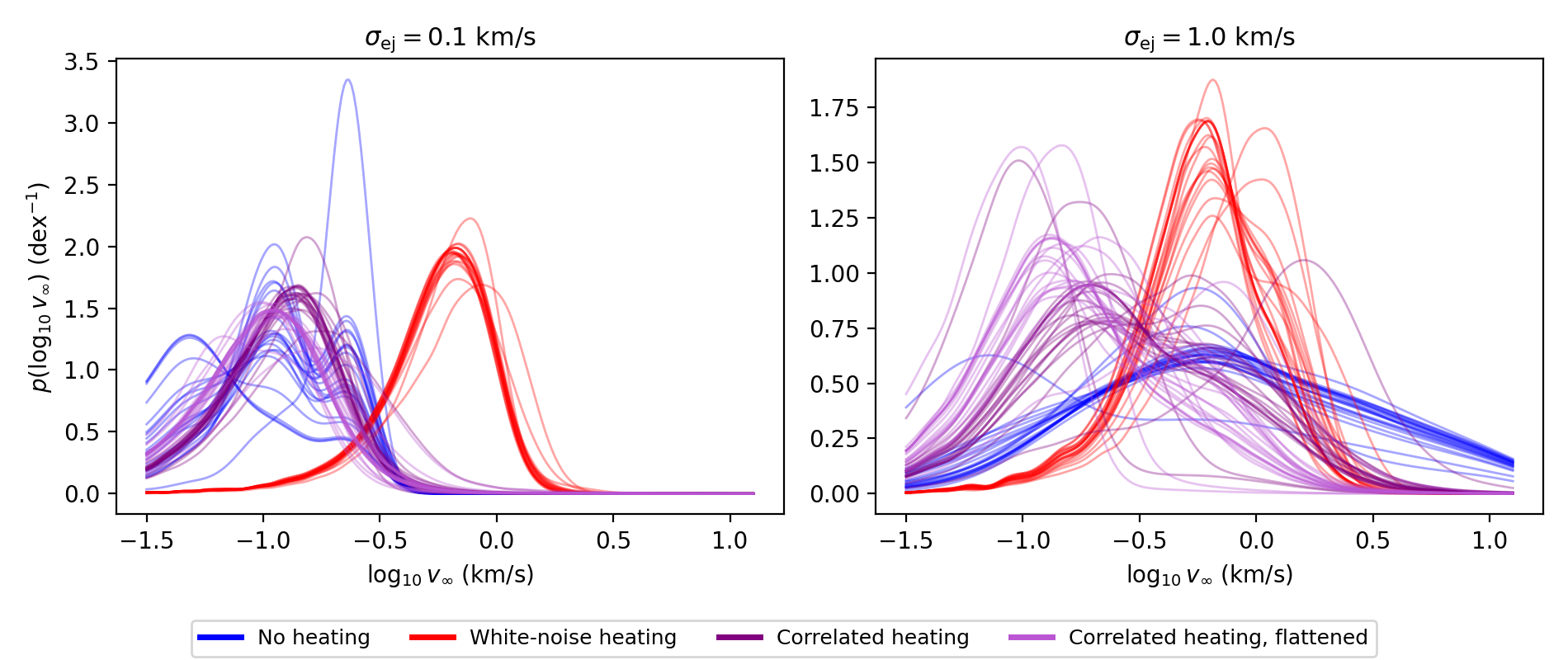}
    \caption{Excess velocity distributions. Each line represents the predicted excess velocity distribution of quasi-ISOs for a particular Oort cloud erosion history, with the different colors representing different heating models. The realistic correlated heating models predict $v_\infty$ values of a few times 0.1 km s$^{-1}$.}
    \label{fig:vdist}
\end{figure*}

In the same way that we can integrate over the return probability as a function of time $\tilde{\mathcal{R}}$, weighted by a given erosion model's loss rate $d N_\mathrm{ej}/dt$ to obtain the total rate $\mathcal{R}$, we can also sum up the distributions of $v_\infty$, the excess velocity of the quasi-ISOs, and $\hat{n}$, their incoming radiants on the sky. For each burst simulation, these two distributions can be estimated using the KDE. When the rate is evaluated (Equation \ref{eq:rtot}), each sample $i$ has a corresponding $\theta$, $\phi$, and $v_\infty$, which are then weighted by that sample's contribution to the rate, i.e. the entire quantity in the $i$th term of the sum. These distributions are then summed together as follows for a given Oort cloud erosion model in analogy to Equation \ref{eq:rtot}:
\begin{equation}
\label{eq:px}
    p(x) \propto \int_0^{t_0} \tilde{\mathcal{R}}(t)\ \frac{dN_\mathrm{ej}}{dt} \left(\sum_k a_k(t) p_k(x) \right).
\end{equation}
Here $p(x)$ is the distribution of some quantity $x$, which may be multi-dimensional, and $k$ indexes the burst simulations for a particular heating and ejection model at a series of different times $t_k$. The $p_k(x)$ is the distribution of quantity $x$ for burst $k$, in practice represented by a set of samples and weights as described above. Finally, $a_k(t)$ represents the interpolation between available burst simulations, so that
\begin{equation}
    a_i(t) =
    \begin{cases}
        \frac{\log t_{i+1} - \log t}{\log t_{i+1} - \log t_i} & t_i\le t<t_{i+1} \\ 
        \frac{\log t - \log t_{i-1}}{\log t_{i} - \log t_{i-1}} & t_{i-1}\le t<t_{i} \\
        1 & t<t_i\ \mathrm{and}\ t_i=\min(\{t_k\}) \\
        1 & t>t_i\ \mathrm{and}\ t_i=\max(\{t_k\})
    \end{cases}
\end{equation}
with the convention that the $t_i$ are sorted in time. Outside of the grid of burst times, the weight just sets the contribution to the distribution to the closest grid point (last two cases).

These weighted sets of samples, additionally weighted by Equation \ref{eq:px}, are then plotted using a simple KDE or healpix histogram respectively. The distribution of $v_\infty$ shows (Figure \ref{fig:vdist}) that, remarkably, for realistic heating models (the purple lines) $v_\infty$ is almost always less than 1 km s$^{-1}$, closer to $\sim 0.2\ \mathrm{km}\ \mathrm{s}^{-1}$. 
This is far less than any of the three known interstellar objects \citep{hopkins2025b}.
In general the predicted distribution of excess velocities for bona fide interstellar objects from other stars has negligible contribution at these low values.

\begin{figure*}
    \centering
    \includegraphics[width=1\linewidth]{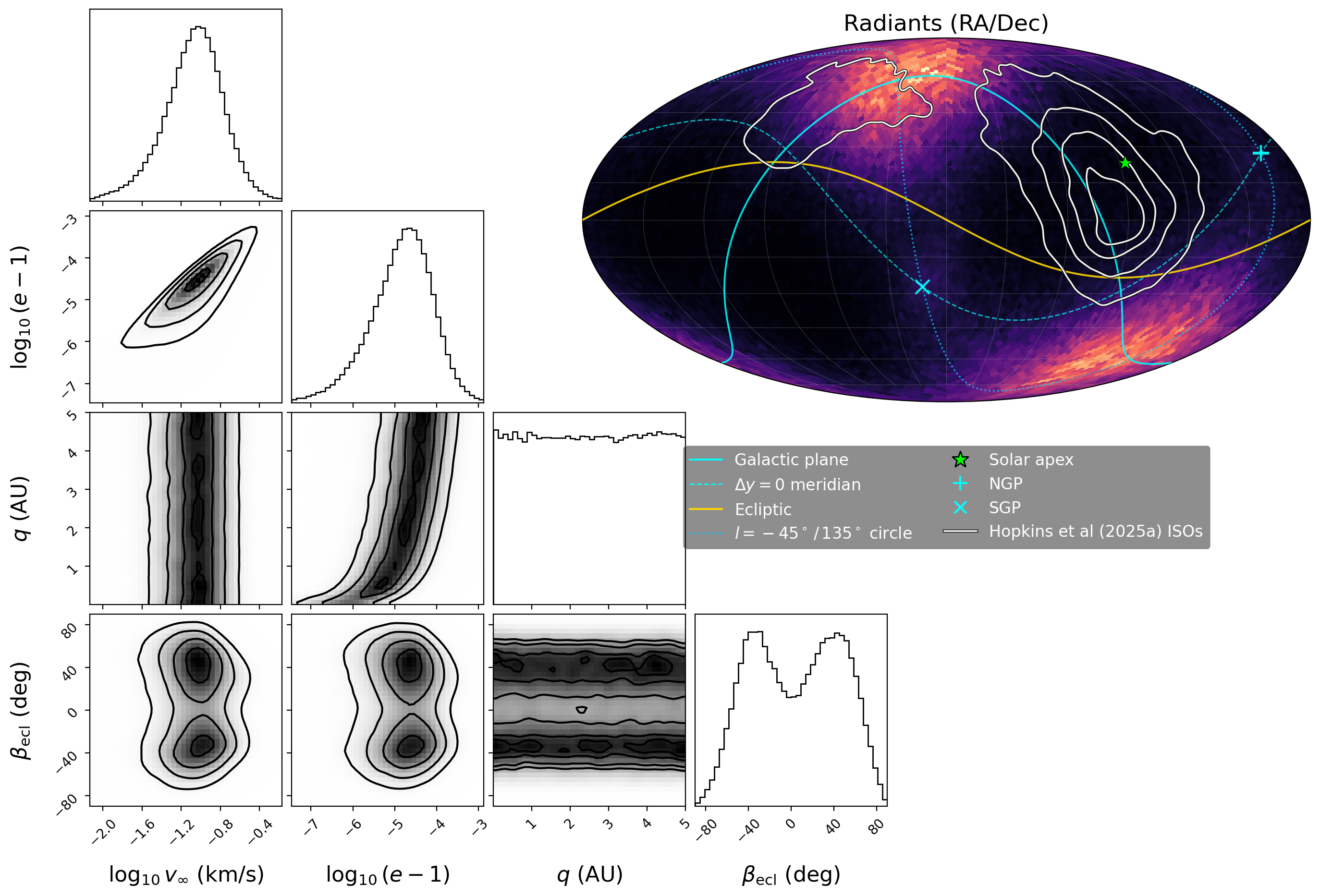}
    \caption{Quasi-ISO properties for the most realistic model: correlated heating with a flattened ejection distribution at $\sigma_\mathrm{ej}=0.1\ \mathrm{km}\ \mathrm{s}^{-1}$ and timing set by the \citet{nesvorny2023} simulation. The quasi-ISOs are dramatically different from the population of ISOs from other stars. The quasi-ISOs have slow excess velocities, eccentricities barely above 1, a gravitational focusing dominated cross-section yielding a constant distribution in pericenter $q$, and a preference for arrival near the Galactic midplane, where it intersects with the $\ell=-45^\circ/+135^\circ$ Galactic meridian. ISOs, shown in the white contours, preferentially arrive from the Solar apex (the direction the Sun is traveling relative to the average motion of nearby stars, shown as the green star) with relatively little overlap with the quasi-ISO population.}
    \label{fig:corner}
\end{figure*}

The on-sky map of quasi-ISO radiants is shown, along with other dynamical properties of the quasi-ISOs, in Figure \ref{fig:corner} for the most realistic model: flattened ejections with correlated Galactic heating, with ejection times set by the \citet{nesvorny2023} simulation. The map is shown in an equatorial coordinate system, with a great circle marking the Galactic midplane, and a great circle representing the $y=0$ meridian, namely objects arriving towards or away from the Galactic center relative to the Sun. For sky maps for other models and Oort cloud ejection timing, see the Appendix. When ejections are spherical, most realizations produce pronounced overdensities at the Galactic poles (see Figure \ref{fig:sky_corr}), connected by a great circle that is not the $y=0$ meridian, but a meridian offset by $\sim 45^\circ$.

The overdensities at the Galactic poles arise from particles that are ejected exactly along the Galactic poles, and therefore have essentially the same in-Galactic-plane epicyclic motion as the Sun. They therefore, relative to the Sun, go straight up and come straight back down, or vice versa. The dynamics are exactly those described in Section \ref{sec:prob}. In other words, not only do these particles preferentially return to the Sun twice per vertical oscillation (where vertical oscillations occur with frequency $\nu$), they do so at a particular incoming radiant, the same one on which they are expelled, close to the Galactic poles.

The meridional offset, in contrast, is somewhat surprising. Particles that are ejected exactly along the $y=0$ meridian are ejected towards or away from the Galactic center as far as their in-Galactic-plane motion is concerned. They then move on their own epicycles similar to the Sun's. These particles have the same $v_\phi$ as the Sun at the time of ejection, but not quite the same angular momentum because of the $dr=1.5$ pc offset in position. If this difference in angular momentum can be neglected, the particle would execute an epicycle relative to the Sun that would return it in the same direction from which it was ejected, i.e. along the $y=0$ meridian in the plots of the radiants. However, the ejection velocities are slow enough that this difference in angular momentum corresponding to a drift velocity, $\sim \Omega dr \sim 0.04\ \mathrm{km}\ \mathrm{s}^{-1}$ cannot be neglected and the particle drifts relative to the Sun over the course of the epicycle. During this time the particle ejected towards the Galactic center orbits slightly faster than the Sun does. Naively this might imply that the radiants should ``lead'' the $y=0$ meridian, i.e. move towards the apex motion of the Sun through the Galaxy, but in fact the incoming radiants trail. This is because the particles returning are not launched with $\Delta y=0$, but rather launched from orbits that would trail the Sun by even more. These two effects combine to yield a great circle passing through $\ell \sim  -45^\circ$. The objects ejected away from the Galactic center similarly trail the Sun's motion but are ejected from a leading population, and populate the other half of the great circle at $\ell \sim +135 ^\circ$.

Putting these two phenomena together, we can understand the distribution of incoming radiants in Figure \ref{fig:corner}. In these simulations, the ejected particles' $z$-velocities in the frame of the Galaxy are reduced by a factor of 10 relative to an isotropic distribution, meaning particles are preferentially ejected in the plane of the Galaxy. The orbits that would preferentially appear at the North and South Galactic Poles are therefore not populated, leaving only particles near the Galactic plane and similarly offset from the $y=0$ great circle by $\sim 45^\circ$.

The corner plot in Figure \ref{fig:corner} also shows that the eccentricities of the quasi-ISOs are above 1 by a very small amount, so that $e-1 ~\sim 10^{-4}$. Additionally, the distribution of pericenters is flat out to 5 au. Both of these are consequences of the very slow excess velocities and the dominance of the gravitational focussing term in the cross-section. Finally, because the quasi-ISOs arrive at two very particular locations, namely the intersection of the Galactic plane with the $\ell = -45^\circ$ meridian, the incoming radiants have an Ecliptic latitude $\beta_\mathrm{ecl}$ that is bimodal with broad peaks around $\pm 50^\circ$.

\section{Discussion}
\label{sec:disc}

\subsection{Comparison to the Interstellar Object Population}
The pressing question around quasi-ISOs is whether they represent a substantial foreground for bona fide ISOs, i.e. when a new ISO is detected, how much do we need to worry that it is actually a quasi-ISO originally from the Solar System? 
Fortunately, we show that quasi-ISOs are dramatically different from bona fide ISOs in essentially every way, which means that mistaking one for the other is very unlikely. 
First, quasi-ISOs are slow, with hyperbolic excess velocities of order 0.1 km s$^{-1}$. 
True ISOs have typical velocities closer to 60 km s$^{-1}$. 
Physically, the small excess velocity of quasi-ISOs is inherited from the small ejection velocity of objects ejected from the Oort cloud by the influence of the Galactic tide and weak stellar flybys. 
This velocity is simply of order the orbital velocity of the outer Oort cloud objects that are unbound in these processes, and is directly visible in the simulation data from \citet{nesvorny2023}: see figure 7 of \citet{albrow2026}. 
While larger velocity ejections can happen for more dramatic flybys \citep[e.g.][]{Pfalzner_2021}, they are both rare, and in any case will not contribute substantially to the quasi-ISO population: the higher velocities will dilute the cloud of ejecta, reducing the chances any of its members will return to the inner Solar System (see Fig.~\ref{fig:probs}). 

In contrast, bona fide ISOs from other stars are expected to have a velocity distribution similar to that of nearby stars, with dispersions of 10s of km s$^{-1}$ in each dimension \citep{hopkins2025a}. This is because they initially inherit the velocity of their parent stars, plus the comparatively small velocities with which they are ejected \citep{albrow2026}, then are subject to the same Galactic dynamical effects and perturbations as the stars. While the \citet{hopkins2025a} model does have a finite low-velocity tail, the fraction of objects with a velocity relative to the Sun $v_\infty < 0.5$ km s$^{-1}$ in the \citet{hopkins2025a} model is small, $1\times 10^{-4}$ (in fact just one star in the sample). 

The second reason to expect quasi-ISOs to be distinct in the observational sample is that quasi-ISOs are likely far rarer than ISOs. 
The models of Oort Cloud erosion that produce the most quasi-ISOs (see Fig.~\ref{fig:erosion_result}) would produce $\sim 2$ quasi-ISOs per decade of 1I/\okina Oumuamua's or larger passing within 5 au of the Sun --- several orders of magnitude below the expected rate for detected ISOs within 5 au \citep{Meech_2017,Do_2018, Dorsey_2025}. 
These maximal quasi-ISO models are the result of assuming that the Oort cloud began its life with $\sim 5\times10^{14}$ objects the size of 1I/\okina Oumuamua or larger. 
While this is possible (see next subsection), this is at the high end of what is plausible. 

Finally, the third way that quasi-ISOs differ from ISOs in their dynamics is that the radiants of ISOs from other stars are preferentially clustered around the Solar Apex, the direction the Sun is moving relative to the local standard of rest \citep{feng2018,hopkins2025a}. Quasi-ISOs meanwhile preferentially avoid that region (see Figures \ref{fig:corner} and the appendix), at least at the present-day position and velocity of the Sun. 
This is the least constraining difference of the three, since the \citet{hopkins2025a} model does include a north-east lobe with some overlap in the quasi-ISO distribution (see Figure \ref{fig:corner}).

\subsection{Normalization Uncertainty}
\label{sec:discnorm}
Our predictions for the absolute number of quasi-ISOs that enter the volume accessible for detection in sky surveys (assumed for simplicity to be a $\sim 5$ au sphere around the Sun) is directly proportional to the assumed normalization in any given Oort cloud erosion model. 
We have adopted for our ``OC Norm'' shown in Fig.\ref{fig:erosion_result} the median value of \citet{boe2019} for objects larger than 100 m in diameter, namely $N_\mathrm{Oort}(t_0) = 5\times 10^{12}$ such objects in the present-day Oort cloud. \citet{boe2019} assign a $0.3 \times10^{12}$ uncertainty on this number in either direction, primarily arising from inferring the fraction of long-period comets that are truly new \citep{festou1993,Wiegert:1999}.
An alternative choice of normalization is that, based on the detection of 1I/\okina Oumuamua in PanSTARRS, each star should contribute on average $N_\mathrm{ISOs\ per\ star}\sim 10^{16}$ objects at least the size of 1I to the ISO population. If the Solar System is typical in this respect and if it promptly ejects the vast majority of its planetesimal population retaining only a small fraction $\epsilon_\mathrm{Oort-form}\sim 5\%$ in the Oort cloud \citep[e.g.][]{dones2004,nesvorny2023}, then we expect the number of Oort cloud objects at the birth of the Oort cloud to be $N_\mathrm{Oort}(0) = \epsilon_\mathrm{Oort-form}N_\mathrm{ISOs\ per\ star} \sim 5\times 10^{14}$. This is our ``ISO Norm'' as shown in Figure \ref{fig:erosion_result}.

Typically these normalizations will yield different values for the number of quasi-ISOs observed per year and any other quantity, e.g. the number of Oort cloud objects today. The Oort cloud is expected to lose mass due to the synergistic effects of stellar flybys and the Galactic tide \citep{rickman2008}, but the rate of these losses is uncertain, mainly owing to the unknown configuration of the Oort cloud itself over time \citep{hanse2018,torres2019}. Values for the ratio $\epsilon_\mathrm{Oort-survival} \equiv N_\mathrm{Oort}(t_0)/N_\mathrm{Oort}(0)$ range from a few percent to about 97\% \citep{portegieszwart2025}, whereas for our two choices of norms to be consistent with each other, we would need $\epsilon_\mathrm{Oort-survival} \approx 5 \times 10^{12}/(5\times 10^{14}) = 10^{-2}$. This is a conceivable but atypical outcome in models of Oort cloud erosion in which $\epsilon_\mathrm{Oort-survival}$ is more like $30\%$, leaving a remaining factor of order 30 discrepancy between the two choices of normalization, visible as the gap between the shaded-in and hollow points in Figure \ref{fig:erosion_result}. Note that the size of the gap varies from model to model owing to the different actual values of $\epsilon_\mathrm{Oort-survival}$ (see e.g. Figure \ref{fig:erosion_models}).

Possibilities for reconciling this discrepancy include the fact that the inner Oort cloud contributes few comets to the effective observable volume of the inner Solar System \citep{hills1981}, so despite efforts to correct for observational biases as in \citet{boe2019}, these objects are likely missed in current surveys. 
LSST will be more sensitive to the higher-pericenter population of inner Oort cloud objects \citep{Ivezic_2019, Inno:2025}, so this uncertainty may be lessened soon. 
There is also some indication that the translation between H band magnitude and physical size may be offset from the typical relation due to difficulties measuring the nucleus \citep{zhao2026}. 
If comet nuclei are larger than previously thought, there is a longer way to extrapolate down the size distribution to 1I/\okina Oumuamua-sized objects, meaning that the \citet{boe2019} value underestimates the present-day value. 
If the small-size end of the size-frequency distribution is as shallow as reported by \citet{boe2019}, this will not be a large correction, but if the slope of the size-frequency distribution is affected by the offset identified in \citet{zhao2026}, the correction could be larger (or smaller). 
Meanwhile on the ISO side, the number is of course highly uncertain because ultimately, despite the subsequent discoveries of 2I/Borisov and 3I/ATLAS, the number density of 1I-sized objects is presently still pinned to a single object and associated search volume \citep{Meech_2017}. However, there are also indications that the discrepancy may be worse, if the current flux of Oort cloud comets has been enhanced by a possible recent flyby of HD 7977 \citep{kaib2026}.

\subsection{Have we seen any quasi-ISOs already?}

The slow excess velocities we predict here for quasi-ISOs mean that they would not currently be identified as ISOs, but near-parabolic Oort cloud objects. Indeed, the population of observed comets with nominally hyperbolic orbits numbers in the hundreds and has an excess velocity distribution not dissimilar from that shown in Figure \ref{fig:vdist} \citep{delafuentemarcos2018}. To distinguish those objects that happen to have hyperbolic osculating orbital elements from those that had hyperbolic orbits before interacting with the planets, one must integrate their orbits backwards to when the object was well outside the planetary region. Of the many objects observed on near-parabolic orbits, some do have barycentric orbital elements corresponding to unbound orbits \citep{Krolikowska2014K}. However, there are substantial systematic uncertainties in estimating the pre-inner-Solar-System orbital elements of these objects due to non-gravitational accelerations \citep{Krolikowska2020} meaning that no single comet in the present catalogue clearly began its passage through the inner Solar System on a hyperbolic orbit. 
Future surveys such as LSST may help here.
Moreover, truly hyperbolic orbits from the Oort cloud can be produced by planetary-mass perturbers \citep{Higuchi_2020}, suggesting that it may be difficult to distinguish quasi-ISOs from this population from excess velocity alone. 
The radiants may offer some disambiguation, since the quasi-ISOs appear to prefer arriving from several distinct directions in the Galactic plane, but this will require further investigation into the properties of the hyperbolic Oort cloud objects.

\subsection{Implications of a future quasi-ISO discovery}

Quasi-ISOs are likely to be rare, and may be difficult to distinguish from the much larger population of near-parabolic Oort cloud objects, therefore we expect the positive identification of a quasi-ISO to be unlikely even with LSST. 

However, the implications of such a discovery are intriguing. 
The main outcome would be that quasi-ISOs are more common than we have estimated here, meaning that the normalizations we have chosen for the Oort cloud are too low, or that a particularly destructive flyby occurred $10-300$ Myr ago. 

In the case of the former, the ``ISO Norm'' shown in Figure \ref{fig:erosion_result} would be favored, and perhaps even too low itself. This is not impossible given the number of objects that may be hiding in the inner Oort cloud \citep[e.g.][]{hills1981,bailey1988}. However, adding a huge quantity of objects to the inner Oort cloud is unlikely to actually change the x-axis values of Figure \ref{fig:erosion_result}, since these objects are much less subject to erosion than the more-loosely bound outer Oort cloud \citep[e.g.][]{torres2019}. We therefore consider this solution unlikely, even in the event of detecting a quasi-ISO.

The other solution, a catastrophic flyby tens to hundreds of Myr ago, is long enough in the past that {\em Gaia} is typically unable to identify good candidates, purely because the star involved would be too faint today to have a precisely-determined orbit \citep{bailerjones2022}. This timescale also begins to be subject to uncertainties in the Galactic potential, including a reconfiguration of the dense gas in the Solar neighborhood \citep[e.g.][]{maconi2025}. The discovery of a quasi-ISO could therefore, pending a much deeper understanding of the local environment and future higher-precision stellar astrometry, represent a unique constraint on the existence of such a flyby.

\section{Conclusion}
\label{sec:conc}

We have assessed the possibility of a new class of small body: quasi-interstellar objects, objects ejected by the Solar System itself that subsequently return to the observable volume. The Solar System produces interstellar objects when the planets or external perturbers (flybys and the Galactic tide) eject planetesimals onto unbound orbits. Just like other interstellar objects, the Sun's interstellar objects orbit the Galaxy in tidal streams, and have the potential to (re-)encounter the Solar System. In fact in general a stream of ISOs is likely to be densest close to its progenitor system \citep{Forbes_2024}. To examine the rate at which we expect quasi-ISOs to appear in the Solar System, we ran a suite of simulations in which the Sun produces a burst of ISOs at various points in time along its orbit. The ISO particles are then integrated in a Galactic potential with one of several prescriptions for dynamical heating, and the rate at which these ISOs re-encounter the Sun is computed. These individual simulation results are convolved with various models of planetesimal ejection from the Solar System, including analytic and Monte Carlo rates of Oort Cloud erosion, as well as simulation results from the long-term Solar System integration of \citet{nesvorny2023}.

We find that the probability of any one randomly-selected ISO from any particular burst re-encountering the Sun each year is $\lesssim 3\times10^{-14}$, depending on the Galactic heating model, the time of the burst in the past, and the properties of the particles at ejection, most notably their speeds. In our most realistic model, the typical ejection velocities are about 0.1 km s$^{-1}$, with ejection directions preferentially concentrated around the Galactic plane, and the particles are then subject to spatial and temporally-correlated heating in the Galaxy. In this model, the bursts that have the greatest probability of supplying a quasi-ISO occur around 60 Myr ago. The probabilities rapidly decrease for bursts older than 1 Gyr, meaning that the huge quantity of planetesimals thought to be lost early in the life of the Solar System do not contribute meaningfully to the quasi-ISO rate. Instead it is the erosion of the Oort cloud in the last $\sim0.5$ Gyr that has a chance of producing quasi-ISOs.

Following convolution of these models with different models for the expected erosion of the Oort cloud, we find that the quasi-ISOs tend to return with hyperbolic excess velocities of about 0.1 km s$^{-1}$ in several particular patches of sky, on the Galactic plane at Galactic longitudes of about $\ell = 45^\circ$ and $\ell=225^\circ$. 

These properties fortunately make quasi-ISOs readily distinguishable from bona fide ISOs. 
Locally, quasi-ISOs require care to distinguish them from the expected population of Solar System comets perturbed onto hyperbolic orbits that happen to intersect the inner Solar System \citep{Higuchi_2020}. 
We are unlikely to have seen a quasi-ISO in present surveys.

However, finding a quasi-ISO in the next decade would be a surprise but by no means impossible. 
In such an event, we may speculate that a flyby eroded a larger-than-expected fraction of the Oort cloud 10-300 Myr ago.

\begin{acknowledgments}

M.T.B. and J.C.F. appreciate support by the Rutherford Discovery Fellowships from New Zealand Government funding, administered by the Royal Society Te Ap\={a}rangi. M.J.H. appreciates support from the Elaine P. Snowden Fellowship.
We thank David Nesvorn\'y for sharing data on the ejected particles from the Solar System simulation described in \citet{nesvorny2023}.
We appreciate discussions with Joe Masiero and the \OO\ group, especially Angus Forrest for his insights into vertical caustics and correlated heating.

\end{acknowledgments}

\software{DifferentialEquations.jl \citep{rackauckas2017adaptive,rackauckas2017differentialequations,rackauckas2020stability}, 
numpy \citep{Harris_2020},
matplotlib \citep{Hunter_2007},
corner \citep{corner},
astropy \citep{AstropyCollaboration_2013,AstropyCollaboration_2018,AstropyCollaboration_2022},
healpix \citep{2005ApJ...622..759G},
healpy \citep{Zonca2019}
          }

\appendix

\section{Sky maps for different realizations}

\begin{figure}
    \centering
    \includegraphics[width=\linewidth]{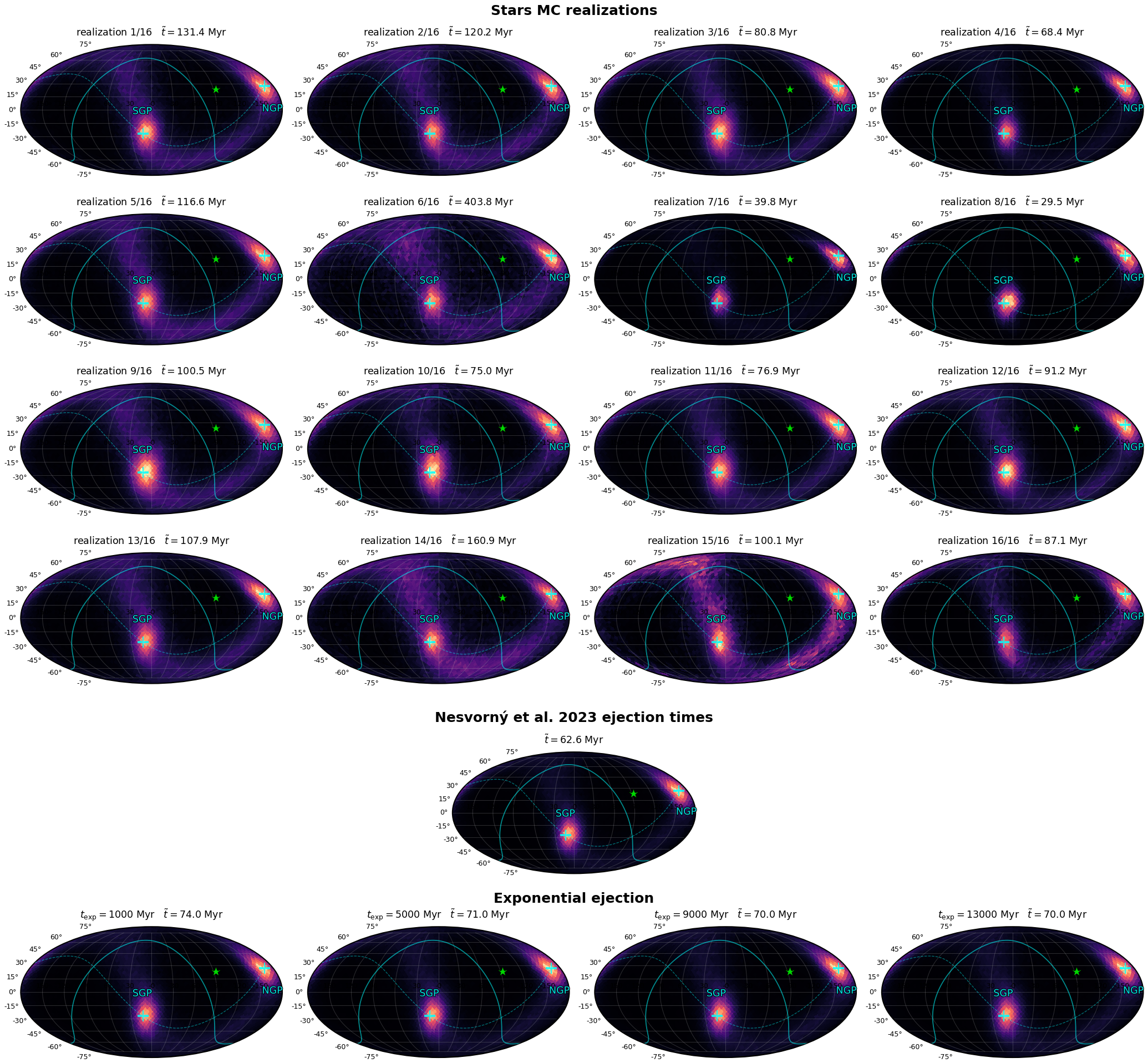}
    \caption{On-sky distribution for the {\em correlated heating} case. Each sky map shows the predicted distribution of incoming radiants on the sky for the quasi-ISOs subject to different Oort cloud erosion models. In each map, we show the North and South Galactic Pole, a great circle in-between them representing the Galactic midplane, and a lighter great circle connecting the poles, namely the $y=0$ meridian. Each panel states the median lookback time of ejection of quasi-ISOs, denoted $\tilde{t}$. The Solar Apex is shown in green.}
    \label{fig:sky_corr}
\end{figure}

\begin{figure}
    \centering
    \includegraphics[width=\linewidth]{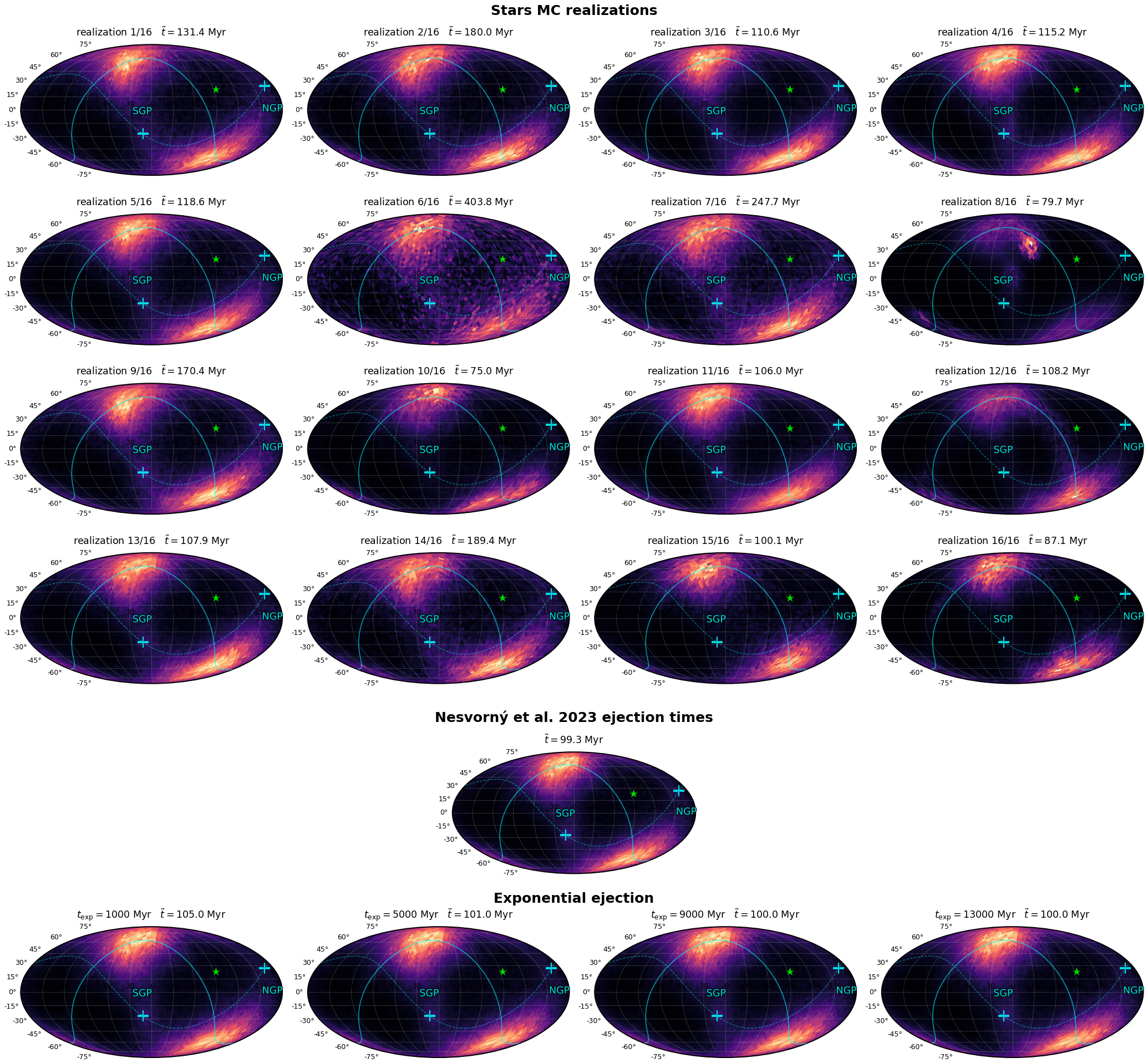}
    \caption{On-sky distribution for the {\em correlated heating with flattened ejection} case. Same as Figure \ref{fig:sky_corr}, but with flattened rather than spherical ejections.}
    \label{fig:sky_corr_compr}
\end{figure}

For each combination of a heating model, ejection model, and Oort cloud erosion model, we can calculate (see section \ref{sec:prop}) an on-sky map of the incoming radiants of the quasi-ISOs. Here we show sky maps for the two correlated heating models: in Figure \ref{fig:sky_corr} the particles are ejected spherically away from the Sun, while in Figure \ref{fig:sky_corr_compr} the particles are preferentially ejected in the Galactic midplane. In both cases, the ejections have a typical velocity of $0.1$ km s$^{-1}$.

\clearpage

\bibliographystyle{aasjournalv7.1}

\bibliography{references}{}

\end{document}